\documentclass[12pt,a4paper]{article}

\usepackage{amsfonts,amsmath,amsthm}
\usepackage{mathrsfs}

\newcommand{\R}{\mathbb{R}}
\newcommand{\C}{\mathbb{C}}
\newcommand{\Z}{\mathbb{Z}}
\newcommand{\N}{\mathbb{N}}
\newcommand{\HH}{\mathcal{H}}

\newcommand{\gZ}{\mathfrak{Z}}
\newcommand{\sH}{\mathscr{H}}

\newtheorem{thm}{Theorem}

\newtheorem{lem}{Lemma}

\theoremstyle{remark}
\newtheorem*{rem*}{Remark}
\newtheorem*{rems*}{Remarks}
\theoremstyle{definition}
\newtheorem*{def*}{Definition}

\newcommand{\Vol}{\mathop\mathrm{Vol}\nolimits}
\newcommand{\supp}{\mathop\mathrm{supp}\nolimits}
\newcommand{\rank}{\mathop\mathrm{rank}\nolimits}
\newcommand{\Ai}{\mathop\mathrm{Ai}\nolimits}
\newcommand{\Bi}{\mathop\mathrm{Bi}\nolimits}

\newcommand{\dd}{\mathrm{d}}
\newcommand{\const}{\mathrm{const}}

\newcounter{point}

\makeatletter
 \newcommand{\lyxaddress}[1]{
   \par {\raggedright #1 
   \vspace{1.4em}
   \noindent\par}
 }
\makeatother

\begin{document}

\title{On a semiclassical formula for non-diagonal matrix elements}

\author{O. Lev, P. \v{S}\v{t}ov\'{i}\v{c}ek}
\date{}

\maketitle

\lyxaddress{Department of Mathematics, Faculty of Nuclear Science,
  Czech Technical University, Trojanova 13, 120 00 Prague, Czech
  Republic\\
  lev@kmlinux.fjfi.cvut.cz, stovicek@kmalpha.fjfi.cvut.cz}

\begin{abstract}
  \noindent
  Let $H(\hbar)=-\hbar^2d^2/dx^2+V(x)$ be a Schr\"odinger operator on
  the real line, $W(x)$ be a bounded observable depending only on the
  coordinate and $k$ be a fixed integer. Suppose that an energy level
  $E$ intersects the potential $V(x)$ in exactly two turning points
  and lies below $V_\infty=\liminf_{|x|\to\infty}\,V(x)$. We consider
  the semiclassical limit $n\to\infty$, $\hbar=\hbar_n\to0$ and
  $E_n=E$ where $E_n$ is the $n$th eigen-energy of $H(\hbar)$. An
  asymptotic formula for $\langle{}n|W(x)|n+k\rangle$, the
  non-diagonal matrix elements of $W(x)$ in the eigenbasis of
  $H(\hbar)$, has been known in the theoretical physics for a long
  time. Here it is proved in a mathematically rigorous manner.
\end{abstract}


\section{Introduction}
\label{sec:introduction}

In the quantum mechanics the matrix elements of an observable occur in
various situations. Let us mention few of them. They measure
transition probabilities between two states and the coefficients in
the stationary perturbation theory are expressed in terms of the
matrix elements of the perturbation. The distribution of matrix
elements is of interest for quantum systems stemming from classically
chaotic systems, see for example
\cite{feingold_peres86:_distrib__matrix_elements_chaotic,
  combescure_robert94:_distrib_matrix_classic_chaotic} and references
in the latter paper. Our immediate motivation to study the matrix
elements was the quantum version of the Kolmogorov-Arnold-Moser method
\cite{bellissard85:_stability_instability_qm},
\cite{duclos_lev_ps_vittot02:_weakly_regular_floquet_hamil}. One of
the assumptions under which this method is applicable is that a
time-dependent perturbation of a quantum system must be sufficiently
small with respect to certain norm which is also expressed in terms of
matrix elements.

One may hope to obtain at least a qualitative information about the
behavior of matrix elements when considering the semiclassical limit.
In fact this idea goes back to the very origins of the quantum
mechanics. A semiclassical formula for non-diagonal matrix elements in
the one-dimensional case has been suggested already a long time ago
\cite{landau_lifshitz58:_quantum_mechanics}. In
\cite{feingold_peres86:_distrib__matrix_elements_chaotic} one can find
another derivation, also on the level of rigor usual in the
theoretical physics, for absolute values of the non-diagonal matrix
elements.

Despite of the ancient history rigorous mathematical results have been
published essentially more recently. Moreover, they cover only some
particular cases even though the technical tools necessary for the
derivation may be at hand nowadays. One usually assumes that the
corresponding classical system is either ergodic
\cite{charbonnel92:_comport_semiclass_ergodiques},
\cite{combescure_robert94:_distrib_matrix_classic_chaotic} or
completely integrable \cite{zelditch90@_quant_transition_amplitudes},
\cite{bellissard_vittot90:_heisenberg_noncommut_semiclass},
\cite{ripamonti96:_class_limit_wigner_fce_bargmann_rep},
\cite{debievre_renaud96:_offdiagonal_matrix_semiclass_limit}. The
semiclassical limit of diagonal matrix elements is now treated in
detail \cite{charbonnel92:_comport_semiclass_ergodiques}. In the case
of multi-dimensional completely integrable systems a formula for
non-diagonal matrix elements was proved in
\cite{zelditch90@_quant_transition_amplitudes},
\cite{ripamonti96:_class_limit_wigner_fce_bargmann_rep},
\cite{debievre_renaud96:_offdiagonal_matrix_semiclass_limit}, see also
\cite{ripamonti98:_class_lobac} for some generalizations. The
one-dimensional case seems to be rather particular. In
\cite{paul_uribe93:_construct_quasimodes_coherent} one can find a
derivation of the semiclassical formula for pseudo-differential
operators in one variable such that the Weyl symbol of the Hamiltonian
is a real polynomial on the phase space while imposing an additional
assumption on the discreteness of the operator spectrum.

The present paper aims to provide a mathematically rigorous
verification of the semiclassical limit of non-diagonal matrix
elements for Schr\"odinger operators on the real line. We prove the
formula under mild assumptions on the potential. In addition, we take
care about identifying the quantum number coming from the
Bohr-Sommerfeld quantization condition with the index determined by
the natural enumeration of eigenvalues in ascending order. Our
approach relies on a transparent application of some well established
tools in the spectral and semiclassical analysis. So we briefly recall
the corresponding results while adjusting their formulation to our
purposes. On the other hand, the chosen method restrict us to
considering observables which depend on the coordinate only. This
particular case was sufficient for the applications we originally had
in mind, as mentioned above.

Let us now formulate precisely in what sense the semiclassical limit
is understood. Set
\begin{equation}
  \label{eq:Hhbar_def}
  H(\hbar) = -\hbar^2\frac{d^2}{dx^2}+V(x)\quad
  \textrm{in~}L^2(\R,dx).
\end{equation}
We consider a fixed energy $E$ and an observable $W=W(x)$ depending
only on the coordinate $x$. The assumptions are as follows.

We suppose that $V(x)$ is bounded from below and three times
continuously differentiable, $W(x)$ is bounded and continuously
differentiable,
\begin{equation}
  \label{eq:def_Vbar}
  E < V_\infty := \liminf_{|x|\to\infty}\,V(x).
\end{equation}
We assume that at the energy $E$ there are exactly two regular turning
points, i.e., $V^{-1}(E)=\{x_-,x_+\}$, $x_-<x_+$, and
$V'(x_\pm)\neq0$. Set
\begin{equation}
  \label{eq:f_def}
  f(x) = V(x)-E.
\end{equation}
In addition we introduce an assumption making it possible to apply the
WKB approximation, namely we assume that
\begin{equation}
  \label{eq:var_F_finite}
  \int_{\R\setminus[-a,a]}\left|\frac{1}{f^{1/4}}\frac{\dd^2}{\dd x^2}
  \!\left(\frac{1}{f^{1/4}}\right)\right|\dd x < \infty
\end{equation}
where $a$ is a positive number chosen so that $f(x)\geq\delta>0$ for
$|x|\geq{}a$. Notice that
\begin{displaymath}
  \frac{1}{f^{1/4}}\frac{\dd^2}{\dd x^2}
  \!\left(\frac{1}{f^{1/4}}\right)
  = \frac{5(V')^2-4(V-E)V''}{16(V-E)^{5/2}}\,.
\end{displaymath}
It may be convenient to replace condition (\ref{eq:var_F_finite}) by
two simpler conditions,
\begin{equation}
  \label{eq:var_F_interms_V}
  \int_{\R\setminus[-a,a]}\frac{|V'|^2}{(V-E)^{5/2}}\,
  \dd x < \infty,\textrm{ }
  \int_{\R\setminus[-a,a]}\frac{|V''|}{(V-E)^{3/2}}\,
  \dd x < \infty.
\end{equation}

The part of the spectrum of $H(\hbar)$ lying below $V_\infty$ is known
to be formed exclusively of simple isolated eigenvalues. We fix the
phase of an eigenfunction $\psi_n$ corresponding to an eigenvalue
$E_n<V_\infty$ by requiring $\psi_n$ to be positive on a neighborhood
of $+\infty$. Moreover, there exists a strictly decreasing sequence of
positive numbers tending to $0$, $\{\hbar_n\}_{n=n_0}^\infty$, and a
constant $\hbar_0>0$ such that for $\hbar\in\,]0,\hbar_0]$, $E$
belongs to the spectrum of $H(\hbar)$ if and only if $\hbar=\hbar_n$
and in that case $E=E_n$ is the $n$th eigenvalue of $H(\hbar)$
provided the enumeration of eigenvalues starts from the index $n=0$.

\textit{Under these assumptions we claim that if $k\in\Z$ is fixed,
  $n\to\infty$, $\hbar=\hbar_n\to0$, with $E=E_n$, then
  \begin{equation}
    \label{eq:nWnk_to_semic}
    \langle{}n|W(x)|n+k\rangle \to \frac{1}{T}
    \int_0^T W\!\left(q(t)\right)e^{ik\omega t}\,\dd t
  \end{equation}
  where $\left(q(t),p(t)\right)$, $t\in[0,T]$, is the classical
  trajectory in the phase space at the energy $E$ and with the initial
  point chosen so that the kinetic energy vanishes, i.e., $p(0)=0$,
  and $q(0)$ coincides the right turning point $x_+$. Furthermore,
  $T>0$ is the period of the classical motion and $\omega=2\pi/T$ is
  the frequency.}

\begin{rem*}
  If the phase of the wave function $\psi_n$ was chosen so that
  $\psi_n$ was positive on a neighborhood of $-\infty$ then formula
  (\ref{eq:nWnk_to_semic}) would be again true with
  $(q(0),p(0))=(x_-,0)$.
\end{rem*}

As already said, we have confined ourselves to observables depending
only on the coordinate because our method of proof is based on the WKB
approximation. One naturally expects, however, that for any smooth
bounded classical observable $A(q,p)$,
\begin{displaymath}
  \langle{}n|\hat{A}|n+k\rangle\to
  \frac{1}{T}\int_0^T A\!\left(q(t),p(t)\right)e^{ik\omega t}\dd t
\end{displaymath}
where $\hat{A}$ is a suitable quantization of $A$. We have already
mentioned that this result is actually proved in
\cite{paul_uribe93:_construct_quasimodes_coherent} in the case when
the potential $V(x)$ is a polynomial.

Let us rewrite the RHS in formula (\ref{eq:nWnk_to_semic}). The
equation of the classical trajectory in the phase space reads
$p^2+V(x)=E$ and its period equals
\begin{equation}
  \label{eq:T_def}
  T=\int_{x_-}^{x_+}\frac{\dd x}{\sqrt{E-V(x)}}\,.
\end{equation}
For $x\in[x_-,x_+]$ set
\begin{equation}
  \label{eq:tau_def}
  \tau(x) = \frac{1}{2}\int_x^{x_+}\frac{\dd y}{\sqrt{E-V(y)}}\,.
\end{equation}
Then $\tau(x_+)=0$, $\tau(x_-)=T/2$, $q(\tau(x))=x$, and
\begin{displaymath}
  \int_0^T W\!\left(q(t)\right)e^{ik\omega t}\,\dd t
  = \int_{x_-}^{x_+}\frac{W(x)}{\sqrt{E-V(x)}}\,
  \cos\!\left(\frac{2\pi k}{T}\,\tau(x)\right)\dd x.
\end{displaymath}

The paper is organized as follows. In Sections 2 through 4 we recall
some preliminaries that we need for the proof of the formula.
Section~2 is devoted to the basic spectral properties of the
Schr\"odinger operator, Section~3 is concerned with the Weyl
asymptotic formula and some basic facts about the WKB approximation
are summarized in Section~4. By counting the zeroes of wave functions
we show in Section~5 that the quantum number coming from the
Bohr-Sommerfeld quantization condition equals the index of the
corresponding eigenvalue. The semiclassical formula is then proved in
Section~6.

\section{Properties of the spectrum lying below $V_\infty$}
\label{sec:spectrum}

Here we briefly recall two well known properties of Schr\"odinger
operators. In the monographs they are usually formulated and derived
for potentials diverging at infinity. We just wish to point up that
the same assertions apply also for more general potentials provided
one takes care only about the part of the spectrum lying below
$V_\infty$. The corresponding proofs can be taken almost literally
from the cited monographs.

In this section (and only in it) the Planck constant is not relevant
and so we set it equal to $1$ and consider the Hamiltonian
\begin{displaymath}
  H = -\frac{\dd^2}{\dd x^2}+V(x)\quad
  \textrm{in }L^2(\R,\dd x).
\end{displaymath}

The following theorem is in fact widely used. We recall it in a form
which is a direct modification of Theorem~XIII.16 in
\cite{reed_simon78:_methods_modern_mf_IV}. Its proof is based on the
min-max principle and is applicable in any dimension of the underlying
Euclidean space. Moreover, the differentiability of $V(x)$ is not
required.

\begin{thm}
  \label{thm:lower_edge_essential}
  Let $V$ be a measurable function in $\R^n$ which is bounded from
  below. Define $H=-\Delta+V$ as the sum of quadratic forms in
  $L^2(\R^n,\dd^nx)$. Then the lower edge of the essential spectrum of
  $H$, if any, is greater than or equal to
  $V_\infty=\liminf_{|x|\to\infty}\,V(x)$.
\end{thm}

Let us note that in the one-dimensional case and provided the
potential is continuous Theorem~\ref{thm:lower_edge_essential} also
follows from a well known estimate on the number of negative
eigenvalues.

Here and everywhere in what follows, if $A$ is a self-adjoint operator
then $P(A;\cdot)$ designates the associated projector-valued measure,
and for $K\in\R$ we denote
\begin{displaymath}
  N(A,K) = \rank{}P(A;]-\infty,K[).
\end{displaymath}
Further, for a real-valued function $W(x)$ we set
\begin{displaymath}
  W_-(x)=\max\{0,-W(x)\}.
\end{displaymath}

It holds (see, for example, Theorem~5.3 in
\cite{berezin_shubin83:_uravnenie_sredingera})
\begin{displaymath}
  N(H,0) \leq 1+\int_{\R}|x|\,V_-(x)\,\dd x.
\end{displaymath}
In particular, if $V(x)$ is continuous and bounded from below then for
any $c<V_\infty$ the function $(V-c)_-(x)$ has a compact support and,
by this estimate, $N(H,c)<\infty$. This again implies that the lower
edge of the essential spectrum of $H$ is greater than or equal to
$V_\infty$.

The next property is specific for the one-dimensional case. The
potential $V(x)$ is supposed to be continuous and bounded from below.

As is well known from the theory of ordinary differential equations,
for $E<V_\infty$, any nontrivial solution of the Schr\"odinger
equation either grows at least exponentially or decays at least
exponentially at $+\infty$ (see, for example, Corollary~1 in
\cite[Section~II]{berezin_shubin83:_uravnenie_sredingera}). The latter
solution is called recessive at $+\infty$ and is unique up to a
multiplicative constant. Of course, an analogous assertion is also
true for $-\infty$. It immediately follows that all eigenvalues of the
Hamiltonian $H$ lying below $V_\infty$ are simple. Moreover, in virtue
of Theorem~\ref{thm:lower_edge_essential}, they have no accumulation
points below $V_\infty$. Consequently, the eigenvalues of $H$ below
$V_\infty$ can be arranged into a strictly increasing sequence, empty
or finite or infinite,
\begin{displaymath}
  E_0<E_1<E_2<\ldots<V_\infty.
\end{displaymath}

The following theorem is a straightforward modification of Theorem~3.5
in \cite[Chapter~II]{berezin_shubin83:_uravnenie_sredingera}.

\begin{thm}
  \label{thm:number_zeroes_nth}
  The number of zeroes of the $m$th eigenfunction of $H$ corresponding
  to the eigenvalue $E_m<V_\infty$ is exactly equal to $m$.
\end{thm}

\section{The Weyl asymptotic formula}
\label{sec:weyl}

In this section we aim to recall the Weyl asymptotic formula
generalized to Schr\"odinger operators. It can be derived from the
Gutzwiller trace formula
\cite{gutzwiller71:_period_orbits_classic_quant_conds} which was
rigorously proved in
\cite{brummelhuis_uribe91:_semiclass_trace_schroed_oper} under the
assumption that the potential is positive and infinitely
differentiable. In \cite{robert98:_semiclass_approx_survey} there is
given a short review of the history and the Weyl asymptotic formula is
recalled even under stricter assumptions which among others mean that
the potential does not grow faster than polynomially. A weaker version
of the formula is also stated in
\cite[Theorem~XIII.79]{reed_simon78:_methods_modern_mf_IV} but only
for compactly supported potentials.

Here we wish to point out that the proof of Theorem~XIII.79 in
\cite{reed_simon78:_methods_modern_mf_IV} can be extended in a
straightforward manner and thus the Weyl asymptotic formula can be
derived just under the assumption that the potential is semi-bounded
and continuous. We restrict ourselves, however, to the one-dimensional
case only. In addition, this approach is quite simple as it is based
merely on an application of the min-max principle and the
Dirichlet-Neumann bracketing. On the other hand, if compared to the
result based on the trace formula, as presented in
\cite{robert98:_semiclass_approx_survey}, the control of the error
term is essentially worse; it is known to be of order $O(1)$ while the
present method only yields the asymptotic behavior of the type
$o(\hbar^{-1})$.

From now on, the Planck constant is again relevant. This means that
the discussion concerns the Hamiltonian $H(\hbar)$ introduced in
(\ref{eq:Hhbar_def}). Since what follows is nothing but a slight
modification of known results we just indicate the basic steps.

First let us recall a definition from
\cite[XIII.15]{reed_simon78:_methods_modern_mf_IV} making it possible
to compare self-adjoint operators defined in different Hilbert spaces.
The symbol $Q(A)$ stands for the form domain of $A$. If
$\psi\in{}Q(A)$ then the scalar product $\langle\psi,A\psi\rangle$ is
automatically understood in the form sense.

\begin{def*}
  Let $\HH_1\subset\HH$ be a closed subspace, let $A$ be a
  semi-bounded self-adjoint operator in $\HH$ and let $B$ be a
  semi-bounded self-adjoint operator in $\HH_1$. We shall write
  $A\leq{}B$ if and only if it holds
  \renewcommand{\labelenumi}{(\roman{enumi})}
  \begin{enumerate}
  \item $Q(A)\supset{}Q(B)$,
  \item $\forall\psi\in{}Q(B)$,
    $\langle\psi,A\psi\rangle\leq\langle\psi,B\psi\rangle$.
  \end{enumerate}
\end{def*}

With the aid of the min-max principle one can show
\cite[XIII.15]{reed_simon78:_methods_modern_mf_IV} that if $A\leq{}B$
then
\renewcommand{\labelenumi}{(\roman{enumi})}
\begin{enumerate}
\item $\forall{}K\in\R$,
  $\rank{}P(A;]-\infty,K[)\geq\rank{}P(B;]-\infty,K[)$,
\item $\forall{}K\in\R$,
  $\rank{}P(A;]-\infty,K])\geq\rank{}P(B;]-\infty,K])$.
\end{enumerate}

The following lemma is analogous to Proposition~2 in
\cite[XIII.15]{reed_simon78:_methods_modern_mf_IV} in the
one-dimensional case and its proof is based on rather elementary
explicit computations of the eigenvalues for the involved operators.

\begin{lem}
  \label{lem:HN_HM_HD_estim}
  Let $I=[a,b]$ be a compact interval. Let us introduce $H_D$, $H_N$
  and $H_M$ as self-adjoint operators in $L^2(I,\dd{}x)$ such that all
  of them act as the differential operator $-\hbar^2\,\dd^2/\dd{}x^2$
  and whose domain is respectively determined by the Dirichlet,
  Neumann and mixed boundary conditions. Then for all $K>0$ it holds
  \renewcommand{\labelenumi}{(\roman{enumi})}
  \begin{displaymath}
    -1 \leq \rank P(H;]-\infty,K[)-\frac{\ell}{\pi\hbar}\sqrt{K}
    \leq \rank P(H;]-\infty,K])-\frac{\ell}{\pi\hbar}\sqrt{K}
    \leq 1,
  \end{displaymath}
  where $H$ is any of the operators $H_D$, $H_N$, $H_M$, and
  $\ell=b-a$ is the length of the interval.
\end{lem}

The following lemma coincides with Proposition~4 in
\cite[XIII.15]{reed_simon78:_methods_modern_mf_IV} in the
one-dimensional case.

\begin{lem}
  \label{lem:H_bracket_DN}
  Let $-\infty<a<b<c<+\infty$ and let $H$ be a self-adjoint operator
  in $L^2([a,c],\dd{}x)$ which acts as the differential operator
  $-\dd^2/\dd{}x^2$ with either the Dirichlet or the Neumann boundary
  condition imposed at each of the points $a$ and $c$ (mixed boundary
  conditions are admitted). Let $H_D^{(1)}$ and $H_N^{(1)}$ be the
  self-adjoint operators in $L^2([a,b],\dd{}x)$ also acting as
  $-\dd^2/\dd{}x^2$ and with the domain being determined by the same
  boundary condition at the point $a$ as imposed in the case of the
  operator $H$ and by the Dirichlet or Neumann boundary condition at
  the point $b$, respectively.  Analogously one introduces the
  self-adjoint operators $H_D^{(2)}$ and $H_N^{(2)}$ in
  $L^2([b,c],\dd{}x)$. Then it holds
  \begin{displaymath}
    H_N^{(1)}\oplus H_N^{(2)} \leq H \leq H_D^{(1)}\oplus H_D^{(2)}.
  \end{displaymath}
\end{lem}

First let us state the Weyl asymptotic formula for a finite interval.
It can be prover in a way very close to the proof of Theorem~XIII.79
in \cite{reed_simon78:_methods_modern_mf_IV}. So we do not reproduce
the proof but let us note that it is based on a limit procedure when
the interval is split into $N$ subintervals of equal length with $N$
tending to $\infty$. In the course of the proof one uses
Lemma~\ref{lem:HN_HM_HD_estim} and \ref{lem:H_bracket_DN}, the
additivity of the numbers $N(A,K)$, i.e.,
\begin{displaymath}
  N(A_1\oplus A_2\oplus\ldots\oplus A_N,K)
  = N(A_1,K)+N(A_2,K)+\ldots+N(A_N,K),
\end{displaymath}
and the fact that the integral on the RHS of (\ref{eq:Weyl_ab}) exists
in the Riemann sense.

\begin{thm}
  \label{thm:limNHK_on_ab}
  Let $-\infty<a<b<+\infty$, $V\in{}C([a,b])$, and let
  \begin{displaymath}
    H_f(\hbar) = -\hbar^2\,\frac{\dd^2}{\dd x^2}+V(x)
  \end{displaymath}
  be a self-adjoint operator in $L^2([a,b],\dd{}x)$ with either the
  Dirichlet or Neumann boundary condition imposed at each of the
  boundary points $a$ and $b$ (mixed boundary conditions are
  admitted). Then for all $K\in\R$,
  \begin{equation}
    \label{eq:Weyl_ab}
    \lim_{\hbar\to0+}\hbar\,N(H_f(\hbar),K)
    = \frac{1}{\pi}\,\int_a^b\sqrt{(V-K)_-(x)}\,\dd x.
  \end{equation}
\end{thm}

Finally let us proceed to the case of the Hamiltonian $H(\hbar)$.

\begin{thm}
  \label{thm:Weyl_Hhbar}
  Let $V\in{}C(\R)$ be a real-valued function which is bounded from
  below. Then for all $K<V_\infty$ it holds true that
  \begin{equation}
    \label{eq:Weyl_Hhbar}
    \lim_{\hbar\to0+}\hbar\,N(H(\hbar),K)
    = \frac{1}{2\pi}\Vol_Z\big(\sH^{-1}(]-\infty,K[)\big)
    = \frac{1}{\pi}\,\int_\R\sqrt{(V-K)_-(x)}\,\dd x
  \end{equation}
  where $\sH(x,p)=p^2+V(x)$ and $\Vol_Z(X)$ designates the Lebesgue
  measure of a measurable set $X$ in the phase space.
\end{thm}

\begin{proof}
  If $K<V_\infty$ then the support of $(V-K)_-$ is compact. Suppose
  that
  \linebreak
  $\supp(V-K)_-\subset[a,b]$, $-\infty<a<b<+\infty$. Set
  \begin{displaymath}
    H_1(\hbar) = -\hbar^2\,\frac{\dd^2}{\dd x^2}-(V-K)_-(x)\quad
    \textrm{in~}L^2(\R,\dd x)
  \end{displaymath}
  and
  \begin{displaymath}
    H_2(\hbar) = -\hbar^2\,\frac{\dd^2}{\dd x^2}+V(x)-K\quad
    \textrm{in~}L^2([a,b],\dd x)
  \end{displaymath}
  with the Dirichlet boundary condition imposed at the points $a$ and
  $b$. Observe that $-(V-K)_-(x)\leq{}V(x)-K$ on $\R$ and so
  $Q(H(\hbar)-K)\subset{}Q(H_1(\hbar))$. Furthermore,
  $L^2([a,b],\dd{}x)$ can be naturally regarded as a subspace in
  $L^2(\R,\dd{}x)$. If $\psi\in{}Q(H_2(\hbar))$ then $\tilde{\psi}$
  defined by $\tilde{\psi}(x)=\psi(x)$ for $x\in[a,b]$,
  $\tilde{\psi}(x)=0$ for $x\in\R\setminus[a,b]$, belongs to
  $Q(H(\hbar)-K)$ ($\tilde\psi$ is an absolutely continuous function).
  This implies that $Q(H_2(\hbar))\subset{}Q(H(\hbar)-K)$. We have
  find that $H_1(\hbar)\leq{}H(\hbar)-K\leq{}H_2(\hbar)$.  Hence
  \begin{displaymath}
    N(H_2(\hbar),0) \leq N(H(\hbar),K) \leq N(H_1(\hbar),0).
  \end{displaymath}
  
  Formula (\ref{eq:Weyl_Hhbar}) for compactly supported potentials is
  stated in
  \cite[Theorem~XIII.79]{reed_simon78:_methods_modern_mf_IV}. Hence it
  holds
  \begin{displaymath}
    \lim_{\hbar\to0+} \hbar\,N(H_1(\hbar),0)
    = \frac{1}{\pi} \int_{\R} \sqrt{(V-K)_-(x)}\,\dd x,
  \end{displaymath}
  and from Theorem~\ref{thm:limNHK_on_ab} we know that
  \begin{displaymath}
    \lim_{\hbar\to0+}\hbar\,N(H_2(\hbar),0)
    = \frac{1}{\pi}\,\int_a^b\sqrt{(V-K)_-(x)}\,\dd x
    = \frac{1}{\pi} \int_{\R} \sqrt{(V-K)_-(x)}\,\dd x.
  \end{displaymath}
  Formula (\ref{eq:Weyl_Hhbar}) for a general potential then follows
  by bracketing.
\end{proof}

For our purposes the following immediate corollary of
Theorem~\ref{thm:Weyl_Hhbar} will be sufficient. Suppose that $V(x)$
is continuously differentiable and an interval $]a,b[$,
$a<b\leq{}V_\infty$, contains at least one regular value of the
classical Hamiltonian $\sH(x,p)$, i.e., there exists
$\lambda\in\,]a,b[$ satisfying $\sH^{-1}(\{\lambda\})\neq\emptyset$
and $V(x)=\lambda$ implies $V'(x)\neq0$. Then the number of
eigenvalues of $H(\hbar)$ in the interval $]a,b[$ tends to infinity as
$\hbar\to0+$.

\section{The WKB method for one and two turning points}
\label{sec:WKB}

Here we summarize some basic facts about the WKB approximation, also
called Liouville-Green approximation, that we need for the proof of
the formula in Section~\ref{sec:derivation}. At the same time we
introduce the necessary notation. We stick to the presentation given
in the monograph \cite{olver74:_asymptotics_special_funct} whose
distinguished feature is that it provides explicit bounds on the error
terms.

Let us first consider the situation with one turning point.
Let $]a,b[\,\subset\R$ be an interval, finite or infinite,
$x_0\in\,]a,b[$, and $f(x)$ be a real-valued function defined on
$]a,b[$ such that $f(x)/(x-x_0)$ is positive and twice continuously
differentiable (hence $f(x_0)=0$, $f'(x_0)>0$). For $x\in\,]a,b[$ set
\begin{subequations}
  \begin{alignat}{2}
    \label{eq:def_zeta_plus}
    \frac{2}{3}\,\zeta^{3/2} &= \int_{x_0}^x\sqrt{f(t)}\,\dd t
    &\quad&\textrm{if~}x\geq x_0, \\
    \label{eq:def_zeta_minus}
    \frac{2}{3}\,(-\zeta)^{3/2} &= \int_{x}^{x_0}\sqrt{-f(t)}\,\dd t
    &&\textrm{if~}x<x_0.
  \end{alignat}
\end{subequations}
Then $\zeta(x)$ is strictly monotone, $\zeta(x)/(x-x_0)$ is positive
and twice continuously differentiable in $]a,b[$, see Lemma~3.1 in
\cite[Chapter~11]{olver74:_asymptotics_special_funct}.

Assume further that
\begin{equation}
  \label{eq:int_x0b_sqrtf}
  \int_{x_0}^b\sqrt{f(t)}\,\dd t = \infty
\end{equation}
and
\begin{eqnarray}
  \label{eq:f_var_finite}
  \int_{]a,b[\,\setminus U_0}\frac{|f''|}{|f|^{3/2}}\,\dd t < \infty,
  \textrm{~}
  \int_{]a,b[\,\setminus U_0}\frac{(f')^2}{|f|^{5/2}}\,\dd t < \infty,
\end{eqnarray}
where $U_0=[x_0-\varepsilon,x_0+\varepsilon]$ and $\varepsilon$ is any
positive number such that $a<x_0-\varepsilon$ and $x_0+\varepsilon<b$.

Notice also that
\begin{equation}
  \label{eq:der_zeta}
  \zeta' = \left(\frac{f}{\zeta}\right)^{\!1/2}\textrm{~and~}
  \zeta'(x_0) = f'(x_0)^{1/3}.
\end{equation}
Denote by $\xi$ the inverse function to $\zeta$. Theorem~3.1 in
\cite[Chapter~11, \S3.3]{olver74:_asymptotics_special_funct} can be
rephrased as follows.
\begin{thm}
  \label{thm:olver_1turn_pt}
  Under the above assumptions, the solution of the differential
  equation
  \begin{equation}
    \label{eq:diff_eq_w_f}
    \hbar^2\,\frac{\dd^2w}{\dd x^2} = f(x)w
  \end{equation}
  which is recessive as $x$ tends to $b$ exists on $]a,b[$, is unique
  up to a multiplicative constant and equals
  \begin{equation}
    \label{eq:olver_psi_Ai}
    \psi(x) = \left(\frac{\zeta}{f}\right)^{\!1/4}
    \bigl(\Ai(\hbar^{-2/3}\zeta)+\varepsilon(\hbar,x)\bigr)
  \end{equation}
  with the error term satisfying the estimates
  \begin{displaymath}
    |\varepsilon(\hbar,x)| \leq \Phi_0(\hbar^{-2/3}\zeta)\,\hbar,
    \textrm{~}
    \left|\frac{\partial\varepsilon(\hbar,x)}{\partial x}\right|
    \leq \left(\frac{f}{\zeta}\right)^{\!1/2}
    \Phi_1(\hbar^{-2/3}\zeta)\,\hbar^{1/3},
  \end{displaymath}
  where $\Phi_0(x)$, $\Phi_1(x)$ are certain continuous positive
  functions on $\R$ such that
  \begin{displaymath}
    \Phi_0(x) \sim
    \begin{cases}
      \displaystyle{
        \const\,\frac{\exp\!\left(-\frac{2}{3}x^{3/2}\right)}{x^{1/4}}}
      & \quad\textrm{as }x\to+\infty, \\
      \noalign{\medskip}
      \displaystyle{
        \const\,\frac{1}{|x|^{1/4}}}
      & \quad\textrm{as }x\to-\infty,
    \end{cases}
  \end{displaymath}
  \begin{displaymath}
    \Phi_1(x) \sim
    \begin{cases}
      \displaystyle{
        \const\,\exp\!\left(-\frac{2}{3}x^{3/2}\right)}
      & \quad\textrm{as }x\to+\infty, \\
      \noalign{\medskip}
      \displaystyle{
        \const}
      & \quad\textrm{as }x\to-\infty.
    \end{cases}
  \end{displaymath}
\end{thm}

Let us now turn to the case when $f(x)$ is given by (\ref{eq:f_def})
and so is defined on the entire real line.  From now on the potential
$V$ satisfies all assumptions as formulated in the Introduction. In
particular, it follows that the function
\begin{equation}
  \label{eq:V_over_xx_minus_xx_plus}
  \frac{V(x)-E}{(x-x_-)(x-x_+)}\textrm{~~is positive on~}\R
  \textrm{~and belongs to~}C^2(\R).
\end{equation}
Moreover, there exists an open neighborhood of $E$, $U_E=\,]E_-,E_+[$,
$E_-<E<E_+$, such that these assumptions apply for any
$\lambda\in\overline{U_E}$ as well.

For $\lambda\in{}U_E$ set
\begin{displaymath}
  \gamma_\lambda = \sH^{-1}(\{\lambda\})
\end{displaymath}
where $\sH(x,p)=p^2+V(x)$. Thus $\gamma_\lambda$ is a closed curve in
the phase space and the energy takes on it the value $\lambda$. Let us
further introduce the action integral,
\begin{equation}
  \label{eq:action_int_gamma}
  J(\lambda) = \int_{\sH(x,p)\leq\lambda}\dd x\dd p
  = \int_{\gamma_\lambda}p\,\dd x
  = 2\int_{x_-(\lambda)}^{x_+(\lambda)}\sqrt{\lambda-V(x)}\,\dd x
\end{equation}
where $x_-(\lambda)<x_+(\lambda)$ are the turning points at the energy
$\lambda$. Then
\begin{equation}
  \label{eq:Tlamda_def}
  T(\lambda) = J'(\lambda)
  = \int_{x_-(\lambda)}^{x_+(\lambda)}
  \frac{\dd x}{\sqrt{\lambda-V(x)}}
\end{equation}
is the period of the classical trajectory in the phase space.

In the following theorem we summarize the result derived in
\cite[Chapter 13, \S8.2]{olver74:_asymptotics_special_funct}.

\begin{thm}
  \label{thm:olver_sequence_hbarn}
  Under the assumptions on $V$ formulated in the Introduction (in
  particular, we assume that condition
  (\ref{eq:V_over_xx_minus_xx_plus}) is fulfilled as well as the
  convergence of the integrals in (\ref{eq:var_F_interms_V})) there
  exist a neighborhood $U_E$ of $E$, $\hbar_0>0$, $n_0\in\N$ and for
  every $\lambda\in{}U_E$ a sequence
  $\{\hbar_n(\lambda)\}_{n=n_0}^\infty$,
  $\hbar_0>\hbar_{n_0}(\lambda)>\hbar_{n_0+1}(\lambda)
  >\hbar_{n_0+2}(\lambda)>\ldots>0$, such that for
  $\hbar\in\,]0,\hbar_0[$ the energy $\lambda$ is an eigenvalue of
  $H(\hbar)$ if and only if $\hbar=\hbar_n(\lambda)$ for some
  $n\geq{}n_0$. Moreover, the sequence $\{\hbar_n(\lambda)\}$
  asymptotically behaves like
  \begin{equation}
    \label{eq:asympt_hbar}
    \hbar_n(\lambda)^{-1} = (2n+1)\pi J(\lambda)^{-1}+O(n^{-1})
  \end{equation}
  where the error term $O(n^{-1})$ decays in $n$ uniformly with
  respect to $\lambda\in{}U_E$.
\end{thm}

\begin{rem*}
  It is known that if $V\in{}C^r(\R)$, with $r\geq1$, and $E<V_\infty$
  is a regular value of $V(x)$ then the action integral $J(\lambda)$
  defined in (\ref{eq:action_int_gamma}) is $r$~times continuously
  differentiable on some neighborhood of $E$ (see, for example,
  \cite{robert98:_semiclass_approx_survey}).
  
  The verification of this assertion is quite elementary in the
  one-dimensional case and with two turning points at the energy $E$.
  For a sufficiently small neighborhood $U_E=\,]E_-,E_+[$ the function
  $V(x)$ is strictly decreasing on the interval $[x_-(E_+),x_-(E_-)]$
  and strictly increasing on $[x_+(E_-),x_+(E_+)]$, with nowhere
  vanishing derivative. Let us write
  \begin{eqnarray*}
    T(\lambda) &=& \left(\int_{x_-(\lambda)}^{x_-(E_-)}
    + \int_{x_-(E_-)}^{x_+(E_-)}
    + \int_{x_+(E_-)}^{x_+(\lambda)}\right)
  \frac{\dd x}{\sqrt{\lambda-V(x)}} \\
  &=& T_-(\lambda)+T_0(\lambda)+T_+(\lambda).
  \end{eqnarray*}
  Clearly, $T_0(\lambda)\in{}C^\infty(U_E)$.  Thus it is sufficient to
  verify that $T_-(\lambda),T_+(\lambda)\in{}C^{r-1}(U_E)$. Let us
  focus only on the latter function. Set
  $W_+=\left(V\big|_{[x_+(E_-),x_+(E_+)]}\right)^{-1}$. Hence $W_+$ is
  $r$~times continuously differentiable. After some elementary
  manipulations one can show that
  \begin{displaymath}
    T_+(\lambda) = \int_{x_+(E_-)}^{x_+(\lambda)}
    \frac{\dd x}{\sqrt{\lambda-V(x)}}
    = 2\sqrt{\lambda-E_-}\int_0^1
    \frac{\dd t}{V'\big(W_+\big(\lambda(1-t^2)+E_-t^2\big)\big)}\,.
  \end{displaymath}
  From the last expression it is obvious that $T_+(\lambda)$ is
  $r-1$~times continuously differentiable.
\end{rem*}

\section{Number of zeroes derived from the WKB method}
\label{sec:num_zeroes}

We need to show that if $\hbar=\hbar_m(\lambda)$ and hence $\lambda$
is an eigenvalue of $H(\hbar)$, as claimed in
Theorem~\ref{thm:olver_sequence_hbarn}, then $\lambda$ is exactly the
$m$th eigenvalue of $H(\hbar)$. According to
Theorem~\ref{thm:number_zeroes_nth}, the index of an eigenvalue lying
below $V_\infty$ equals the number of zeroes of the corresponding
eigenfunction. Fortunately, the WKB approximation, as explained in
\cite{olver74:_asymptotics_special_funct}, is precise enough to
control the number of zeroes.

Let us recall some facts concerning the Airy functions. Let us denote
by $a_n$ and $b_n$ the zeroes of the Airy functions $\Ai(x)$ and
$\Bi(x)$, respectively, arranged in ascending order of the absolute
value, i.e., $\ldots<b_3<a_2<b_2<a_1<b_1<0$. It is known that
\begin{equation}
  \label{eq:root_AiBi}
  a_n = -\left(\frac{3}{2}\,\pi\!\left(n-\frac{1}{4}\right)
  +\gZ\!\left(n-\frac{1}{4}\right)\right)^{\!2/3}\!,\textrm{~~}
  b_n = -\left(\frac{3}{2}\,\pi\!\left(n-\frac{3}{4}\right)
  +\gZ\!\left(n-\frac{3}{4}\right)\right)^{\!2/3}\!,
\end{equation}
where $\gZ(x)=O(x^{-1})$.

First we again consider the situation with one turning point.  Recall
defining relations (\ref{eq:def_zeta_plus}), (\ref{eq:def_zeta_minus})
for $\zeta$.  In the following theorem we summarize the results from
\S\S~6.1, 6.2 and 6.3 in
\cite[Chapter~11]{olver74:_asymptotics_special_funct}.

\begin{thm}
  \label{thm:WKB_num_zeroes}
  Under the same assumptions as in Theorem~\ref{thm:olver_1turn_pt},
  let $w(x)$ be a nonzero solution of the differential equation
  (\ref{eq:diff_eq_w_f}) on $]a,b[$ which is recessive as $x$ tends to
  $b$ (hence $w(x)$ is unique up to a multiplicative constant). Then
  the set of zeroes of $w(x)$ in $]a,b[$, denoted $\{z_n\}_{n\geq1}$
  and arranged in descending order, is at most countable. Any such a
  zero $z$ fulfills $\zeta(z)<\hbar^{2/3}b_1$.  Furthermore, for all
  sufficiently small $\hbar$ it is true that if
  $\zeta(a)<\hbar^{2/3}b_{n+1}$ then the $n$th zero, $z_n$, does exist
  and obeys the estimate
  \begin{displaymath}
    \hbar^{2/3}b_{n+1} < \zeta(z_n) < \hbar^{2/3}b_n.
  \end{displaymath}
  Moreover,
  it holds
  \begin{displaymath}
    |\zeta(z_n)-\hbar^{2/3}a_n| = O(n^{-1/3})\hbar
  \end{displaymath}
  where the symbol $O(n^{-1/3})$ is uniform with respect to $\hbar$.
\end{thm}

\begin{rems*}
  From Theorem~\ref{thm:WKB_num_zeroes} it immediately follows that
  there are no zeroes in the interval $[x_0,b[$. Furthermore, the
  number of zeroes of $w(x)$ in any fixed nonempty subinterval
  $]c,d[\,\subset\,]a,x_0[$ tends to infinity as $\hbar\to0+$.
\end{rems*}

Now we come back to the case when $f(x)$ is given by (\ref{eq:f_def}),
with $V(x)$ satisfying the assumptions from the Introduction. In
particular, there are two turning points at the energy $E$, $x_-$ and
$x_+$, and $V(x)$ satisfies (\ref{eq:V_over_xx_minus_xx_plus}) and
(\ref{eq:var_F_interms_V}). Then for any $a$, $x_-<a<x_+$, the
function $f(x)$ satisfies the assumptions of
Theorem~\ref{thm:WKB_num_zeroes} with $b=+\infty$ and $x_0$ being
replaced by $x_+$. Actually, condition (\ref{eq:var_F_interms_V})
implies (\ref{eq:f_var_finite}) and condition (\ref{eq:int_x0b_sqrtf})
is fulfilled automatically for $E<V_\infty$. Analogous arguments apply
also for the other turning point $x_-$.

According to Theorem~\ref{thm:olver_sequence_hbarn} there exist
$\hbar_0>0$ and a sequence $\{\hbar_n\}_{n=n_0}^\infty$,
$\hbar_0>\hbar_{n_0}>\hbar_{n_0+1}>\hbar_{n_0+2}>\ldots>0$, such that
for $\hbar\in\,]0,\hbar_0[$, $E$ is an eigenvalue of $H(\hbar)$ if and
only $\hbar=\hbar_n$ for some $n\geq{}n_0$. Let $\psi_n(x)$ be an
eigenfunction of $H(\hbar_n)$ corresponding to the eigenvalue $E$.
Thus $\psi_n(x)$ is recessive both at $+\infty$ and $-\infty$ and is
unique up to a multiplicative constant. We can suppose that $\hbar_0$
is sufficiently small so that $\psi_n(x)$ has at least one zero in the
interval $]x_-,x_+[$. By Theorem~\ref{thm:WKB_num_zeroes}, $\psi_n(x)$
has no zeroes in the set $\R\setminus\,]x_-,x_+[$.

Let us choose a point $x_1\in\,]x_-,x_+[$ independently of $n$. Let
$x_1'$ be the zero of $\psi_n$ which is nearest to $x_1$. This means
that $x_1'$ depends on $n$ but the distance between $x_1$ and $x_1'$
tends to zero as $n$ tends to infinity. Denote by $m_+$ and $m_-$ the
number of zeroes of $\psi_n$ in the interval $[x_1',x_+[$ and
$]x_-,x_1']$, respectively (hence the zero $x_1'$ is counted both in
$m_+$ and $m_-$). Denote by $\zeta_+(x)$ the function defined by
relations (\ref{eq:def_zeta_plus}) and (\ref{eq:def_zeta_minus}), with
$x_0$ being replaced by $x_+$.  In virtue of
Theorem~\ref{thm:WKB_num_zeroes}, there exists a constant $c_+\geq0$
(independent of $n$) such that
\begin{displaymath}
  |\zeta_+(x_1')-\hbar_n^{\,2/3}a_{m_+}|
  \leq \frac{c_+\hbar_n}{m_+^{\,1/3}}
\end{displaymath}
for all $n\geq{}n_0$. An application of the mean value theorem,
\begin{displaymath}
  |u^{3/2}-v^{3/2}| \leq \frac{3}{2}\left(\max\{u,v\}\right)^{1/2}
  |u-v|\textrm{~~for~}u>0,v>0,
\end{displaymath}
yields the inequality
\begin{equation}
  \label{eq:zetan_am_plus}
  \bigl||\zeta_+(x_1')|^{3/2}-\hbar_n |a_{m_+}|^{3/2}\bigr|
  \leq \frac{3}{2}
  \left(\frac{3}{2}\int_{x_-}^{x_+}\sqrt{E-V(x)}\,\dd x\right)^{\!1/3}
  \frac{c_+\hbar_n}{m_+^{\,1/3}}
\end{equation}
which is valid for all sufficiently large $n$. Analogously, for the
other turning point we get the estimate
\begin{equation}
  \label{eq:zetan_am_minus}
  \bigl||\zeta_-(x_1')|^{3/2}-\hbar_n |a_{m_-}|^{3/2}\bigr|
  \leq \frac{3}{2}
  \left(\frac{3}{2}\int_{x_-}^{x_+}\sqrt{E-V(x)}\,\dd x\right)^{\!1/3}
  \frac{c_-\hbar_n}{m_-^{\,1/3}}
\end{equation}
where again $c_-\geq0$ is a constant independent of $n$. Set
\begin{displaymath}
  c = \left(\frac{3}{2}\int_{x_-}^{x_+}\sqrt{E-V(x)}\,\dd x
  \right)^{\!1/3}\max\{c_-,c_+\}.
\end{displaymath}
Combining (\ref{eq:zetan_am_plus}) and (\ref{eq:zetan_am_minus}) we
arrive at the inequality
\begin{displaymath}
  \left|\frac{1}{\hbar_n}\int_{x_-}^{x_+}\sqrt{E-V(x)}\,\dd x
    -\frac{2}{3}\left(|a_{m_-}|^{3/2}+|a_{m_+}|^{3/2}\right)\right|
  \leq c\left(\frac{1}{m_-^{\,1/3}}+\frac{1}{m_+^{\,1/3}}\right).
\end{displaymath}

Let $m=m(n)$ be the number of zeroes of $\psi_n(x)$. Obviously,
$m=m_-+m_+-1$. Recalling the asymptotic behavior of $\hbar_n$, as
stated in (\ref{eq:asympt_hbar}) (see also
(\ref{eq:action_int_gamma})), as well as the asymptotic formulas
(\ref{eq:root_AiBi}) for the roots of the Airy functions we finally
find that
\begin{displaymath}
  \left|n-m+O(n^{-1})-\gZ\!\left(m_--\frac{1}{4}\right)
    -\gZ\!\left(m_+-\frac{1}{4}\right)\right|
  \leq \frac{c}{\pi}
  \left(\frac{1}{m_-^{\,1/3}}+\frac{1}{m_+^{\,1/3}}\right).
\end{displaymath}
By Theorem~\ref{thm:WKB_num_zeroes}, both $m_-$ and $m_+$ tend to
infinity as $n$ tends to infinity. This implies that $m(n)=n$ for all
sufficiently large $n$ and therefore, in virtue of
Theorem~\ref{thm:number_zeroes_nth}, $E$ is the $n$th eigenvalue of
the Hamiltonian $H(\hbar_n)$ (with the numbering starting from $n=0$).

All estimates can be carried out in a uniform manner for $E$ being
replaced by $\lambda$ running over some neighborhood of $E$. We
conclude that

\textit{with the assumptions on $V(x)$ formulated in the Introduction,
  there exist $n_0\in\N$ and a neighborhood $U_E$ of $E$ such that for
  all $n\geq{}n_0$ and $\lambda\in{}U_E$, $\lambda$ equals exactly the
  $n$th eigenvalue of $H\bigl(\hbar_n(\lambda)\bigr)$ (with
  $\hbar_n(\lambda)$ introduced in
  Theorem~\ref{thm:olver_sequence_hbarn})}.

\section{Proof of the formula}
\label{sec:derivation}

Here we prove the limit (\ref{eq:nWnk_to_semic}). We know that there
exists a sequence of positive numbers, $\{\hbar_n\}_{n=n_0}^\infty$,
such that $E$ is the $n$th eigenvalue of $H(\hbar_n)$
(Theorem~\ref{thm:olver_sequence_hbarn}). This sequence is strictly
decreasing and tends to 0. We even known that $\hbar_n\sim{}n^{-1}$ as
$n\to\infty$ (see (\ref{eq:asympt_hbar})). Therefore everywhere in
what follows the symbol $O(\hbar)$ should be understood as a
substitute for $O(n^{-1})$.

Let us fix $x_1,x_1',x_1''\in\,]x_-,x_+[$, $x_1'<x_1<x_1''$. For a
given $\hbar=\hbar_n$ we shall denote by $\psi$ a conveniently
normalized eigenfunction corresponding to the eigenvalue $E=E_n$.
Hence $\psi$ is recessive both at $+\infty$ and $-\infty$. The
normalization is fixed by requiring the eigenfunction $\psi$ to
coincide on the interval $]x_1',+\infty[$ with the solution described
in Theorem~\ref{thm:olver_1turn_pt} (with $f(x)=V(x)-E$ and $x_0=x_+$
being the single turning point in this interval).
Theorem~\ref{thm:olver_1turn_pt} is also applicable to the interval
$]-\infty,x_1''[$ containing the turning point $x_-$. On this
interval, $\psi$ equals $\kappa$ times the solution described in
Theorem~\ref{thm:olver_1turn_pt} for some $\kappa\in\C\setminus\{0\}$.

There exists a neighborhood of $E$, $U_E=\,]E_-,E_+[$, such that any
$\lambda\in{}U_E$ satisfies the same assumptions as those imposed on
$E$. Recall that we have fixed $k\in\Z$. For all sufficiently large
$n$, the $(n+k)$th eigenvalue of $H(\hbar_n)$, called $E_{n+k}$,
exists and lies in $U_E$. For brevity we shall denote $E_{n+k}$
sometimes by $\tilde{E}$. We show below that $\tilde{E}-E=O(\hbar)$,
see (\ref{eq:Eprim-E_eq_Oh}). The eigenfunction of $H(\hbar_n)$
corresponding to the eigenvalue $\tilde{E}=E_{n+k}$ and coinciding on
$]x_1',+\infty[$ with the solution from
Theorem~\ref{thm:olver_1turn_pt} will be denoted by $\tilde{\psi}$. In
this case, too, there exists $\tilde\kappa\in\C\setminus\{0\}$ such
that on the interval $]-\infty,x_1''[\,$, $\tilde\psi$ equals
$\tilde\kappa$ times the solution from
Theorem~\ref{thm:olver_1turn_pt}. Furthermore, denote by
$\tilde{x}_\pm$ the turning points corresponding to $\tilde{E}$, i.e.,
$V(\tilde{x}_\pm)=\tilde{E}$. Since
$V(\tilde{x}_\pm)-V(x_\pm)=\tilde{E}-E$ and $V'(x_\pm)\neq0$ it is
clear that $\tilde{x}_\pm-x_\pm=O(\hbar)$ as well.

The verification of (\ref{eq:nWnk_to_semic}) is based on a series of
estimates relying on Theorem~\ref{thm:olver_1turn_pt}. This will be
done in several steps.

\setcounter{point}{0}

\noindent\addtocounter{point}{1}\textit{(\thepoint)~Relation between
  $\tilde{E}$ and $E$}.
Let $E_m(\hbar)$ be the $m$th eigenvalue of $H(\hbar)$. From the
perturbation theory \cite{kato66:_perturb_theory} one deduces that if
it exists and lies below $V_\infty$ then $E_m(\hbar)$ is strictly
increasing and real analytic as a function of $\hbar$. According to
the conclusion of Section~\ref{sec:num_zeroes}, $E_m(\hbar)$ and
$\hbar_m(\lambda)$ are mutually inverse functions. Therefore if
$\hbar=\hbar_n(E)$ then $\hbar=\hbar_{n+k}(\tilde{E})$. Thus we have
\begin{displaymath}
  \hbar_n(E) = \hbar_{n+k}(\tilde{E})
\end{displaymath}
and from the asymptotic formula (\ref{eq:asympt_hbar}) we get
\begin{displaymath}
  (2n+2k+1)J(E)-(2n+1)J(\tilde{E}) = O(n^{-1}).
\end{displaymath}
Since
\begin{displaymath}
  J(\tilde{E}) = J(E)
  +\frac{\partial J(E)}{\partial\lambda}(\tilde{E}-E)+O((E'-E)^2)
\end{displaymath}
we finally arrive at the equation
\begin{displaymath}
  \frac{2k}{2n+1}\,\frac{J(E)}{T(E)}-\tilde{E}+E
  = O(n^{-2})+O((\tilde{E}-E)^2)
\end{displaymath}
whose solution satisfies
\begin{equation}
  \label{eq:Eprim-E_eq_Oh}
  \tilde{E} = E+\frac{J(E)}{T(E)}\,\frac{k}{n}+O(n^{-2}).
\end{equation}

\noindent\addtocounter{point}{1}\textit{(\thepoint)~Asymptotic behavior of
  $\kappa$ and $\tilde\kappa$}.
On the interval $]x_1',x_1''[$ one can compare the asymptotics of the
solutions which are respectively recessive at $+\infty$ and $-\infty$
and infer this way the asymptotic behavior of $\kappa$ as $\hbar\to0$.
For a moment we shall distinguish by a subscript the functions
$\zeta_\pm$ related to the turning points $x_\pm$ and defined
respectively on the intervals $[x_1',+\infty[$ and $]-\infty,x_1'']$.
Thus
\begin{displaymath}
  \frac{2}{3}\,|\zeta_+|^{2/3} = \left|\int_{x_+}^x
    |f(t)|\,\dd t\right|,\textrm{~~}
  \frac{2}{3}\,|\zeta_-|^{2/3} = \left|\int_x^{x_-}
    |f(t)|\,\dd t\right|,
\end{displaymath}
and both $\zeta_+/f$ and $\zeta_-/f$ are positive functions on their
domains. We have
\begin{displaymath}
  \psi(x) = \left(\frac{\zeta_+}{f}\right)^{\!1/4}
  \left(\Ai(\hbar^{-2/3}\zeta_+)+\varepsilon_+(\hbar,x)\right)
\end{displaymath}
for $x\geq{}x_1'$, and
\begin{displaymath}
  \psi(x) = \kappa\left(\frac{\zeta_-}{f}\right)^{\!1/4}
  \left(\Ai(\hbar^{-2/3}\zeta_-)+\varepsilon_-(\hbar,x)\right)
\end{displaymath}
for $x\leq{}x_1''$. Suppose that $x\in[x_1',x_1'']$. Recalling that
\begin{equation}
  \label{eq:asympt_Ai_minf}
  \Ai(-z) = \frac{1}{\pi^{1/2}z^{1/4}}\left(
    \cos\!\left(\frac{2}{3}\,z^{3/2}-\frac{\pi}{4}\right)
  +O(z^{-3/2})\right)\textrm{~~as~}z\to+\infty
\end{equation}
and the error term estimates from Theorem~\ref{thm:olver_1turn_pt} we
arrive at the equality
\begin{displaymath}
  \cos\!\left(\frac{2}{3}\,\hbar^{-1}|\zeta_+|^{3/2}
    -\frac{\pi}{4}\right) + O(\hbar)
  = \kappa\left(\cos\!\left(\frac{2}{3}\,\hbar^{-1}|\zeta_-|^{3/2}
      -\frac{\pi}{4}\right) + O(\hbar)\right).
\end{displaymath}
Furthermore, in virtue of (\ref{eq:asympt_hbar}) it holds
\begin{displaymath}
  \frac{2}{3}\,\hbar^{-1}\left(|\zeta_+|^{2/3}+|\zeta_-|^{2/3}\right)
  = \hbar^{-1}\int_{x_-}^{x_+}|f(t)|\,\dd t
  =  \left(n+\frac{1}{2}\right)\pi + O(\hbar).
\end{displaymath}
Combining the last two equalities we find that
\begin{displaymath}
  \cos\!\left(\frac{2}{3}\,\hbar^{-1}|\zeta_+|^{3/2}
    -\frac{\pi}{4}\right) + O(\hbar)
  = \kappa\left((-1)^n\cos\!\left(
      \frac{2}{3}\,\hbar^{-1}|\zeta_+|^{3/2}
      -\frac{\pi}{4}\right) + O(\hbar)\right).
\end{displaymath}
For $\hbar$ sufficiently small it clearly exists $x\in[x_1',x_1'']$
such that
\begin{displaymath}
  \cos\!\left(\frac{2}{3}\,\hbar^{-1}|\zeta_+|^{3/2}
    -\frac{\pi}{4}\right) = 1.
\end{displaymath}
It follows immediately that
\begin{equation}
  \label{eq:kappa_asympt}
  \kappa = (-1)^n+O(\hbar).
\end{equation}
Similarly,
\begin{equation}
  \label{eq:tkappa_asympt}
  \tilde\kappa = (-1)^{n+k}+O(\hbar).
\end{equation}

\noindent\addtocounter{point}{1}\textit{(\thepoint)~The leading
  asymptotic term on the interval $]x_+-\delta,\infty[$}.
Fix $\delta>0$ sufficiently small (at least $x_1<x_+-\delta$). Let us
show that
\begin{equation}
  \label{eq:int_psi2_infty}
  \int_{x_+-\delta}^\infty \psi^2\,\dd x = \delta^{1/2}O(\hbar^{1/3}),
  \textrm{~}
  \int_{-\infty}^{x_-+\delta} \psi^2\,\dd x = \delta^{1/2}O(\hbar^{1/3}).
\end{equation}
We shall verify only the first equality in (\ref{eq:int_psi2_infty}).
In view of (\ref{eq:kappa_asympt}) and (\ref{eq:tkappa_asympt}), the
verification of the second one is analogous.

Here and everywhere in what follows the symbol $O(\hbar^\varepsilon)$
should be interpreted properly. It means that there exists a constant
$c\geq0$ (independent of $\delta$) and $\hbar_0(\delta)>0$ such that
for all $\hbar$, $0<\hbar<\hbar_0(\delta)$, it holds
$|O(\hbar^\varepsilon)|\leq{}c\hbar^\varepsilon$.

First let us estimate the contribution from the leading asymptotic
term of $\psi$. Applying the substitution $x=\xi(\hbar^{2/3}z)$ we get
the expression
\begin{equation}
  \label{eq:int_Ai2_subst}
  \int_{x_+-\delta}^\infty \left(\frac{\zeta}{f}\right)^{\!1/2}
  \Ai(\hbar^{-2/3}\zeta)^2\,\dd x
  = \hbar^{4/3}\int_{\hbar^{-2/3}\zeta(x_+-\delta)}^\infty
  \frac{z}{f\bigl(\xi(\hbar^{2/3}z)\bigr)}\,\Ai(z)^2\,\dd z.
\end{equation}
By the assumptions, there exist $x_2>x_+$ and $c_1>0$ such that
$f(x)\geq{}c_1$ for $x\geq{}x_2$. The function $\zeta(x)/f(x)$ is
continuous on the interval $[x_1,x_2]$ and therefore it is majorized
on this interval by a constant $c_2\geq0$. This also means that
\begin{displaymath}
  0<\frac{y}{f\bigl(\xi(y)\bigr)}\leq c_2
  \textrm{~~for~}\zeta(x_1)\leq y\leq\zeta(x_2).
\end{displaymath}
This way we get the following upper bound on (\ref{eq:int_Ai2_subst}),
namely
\begin{eqnarray*}
  && \hbar^{2/3}
  \int_{\hbar^{-2/3}\zeta(x_+-\delta)}^{\hbar^{-2/3}\zeta(x_2)}
  c_2\Ai(z)^2\,\dd z
  + \hbar^{4/3}\int_{\hbar^{-2/3}\zeta(x_2)}^\infty
  \frac{z}{c_1}\,\Ai(z)^2\,\dd z \\
  && \leq\, c_2\hbar^{2/3}\bigl(\Ai'(x)^2-x\Ai(x)^2\bigr)
  \Big|_{x=\hbar^{-2/3}\zeta(x_+-\delta)} + o(\hbar^{4/3}).
\end{eqnarray*}
Here we have used the knowledge of the primitive function
\begin{displaymath}
  \int\Ai(x)^2\,\dd{}x=x\Ai(x)^2-\Ai'(x)^2.
\end{displaymath}
In addition to formula (\ref{eq:asympt_Ai_minf}) let us recall also
the asymptotic behavior of the derivative of the Airy function,
\begin{equation}
  \label{eq:asympt_Aider_minf}
  \Ai'(-z) = \frac{z^{1/4}}{\pi^{1/2}}\left(
    \sin\!\left(\frac{2}{3}\,z^{3/2}-\frac{\pi}{4}\right)
  +O(z^{-3/2})\right)\textrm{~~as~}z\to+\infty.
\end{equation}
Since $\zeta(x_+-\delta)=-\zeta'(y)\delta$ for some
$y\in[x_+-\delta,x_+]$ we find that for
$x=\hbar^{-2/3}\zeta(x_+-\delta)$ it holds
\begin{displaymath}
  |\hbar^{2/3}\Ai'(x)^2|
  \leq \const\,\hbar^{2/3}\left(\hbar^{-2/3}\delta\right)^{\!1/2}
  = \const\,\hbar^{1/3}\delta^{1/2}
\end{displaymath}
and
\begin{displaymath}
  |\hbar^{2/3}x\Ai(x)^2|
  \leq \const\,\hbar^{2/3}\hbar^{-2/3}\delta
  \left(\hbar^{-2/3}\delta\right)^{\!-1/2}
  = \const\,\hbar^{1/3}\delta^{1/2}.
\end{displaymath}
We have shown that
\begin{displaymath}
  \int_{x_+-\delta}^\infty \left(\frac{\zeta}{f}\right)^{\!1/2}
  \Ai(\hbar^{-2/3}\zeta)^2\,\dd x
  = \delta^{1/2}O(\hbar^{1/3}).
\end{displaymath}

\noindent\addtocounter{point}{1}\textit{(\thepoint)~The error
  term on the interval $]x_+-\delta,\infty[$}.
Further let us write
\begin{displaymath}
  \psi^2 = \left(\frac{\zeta}{f}\right)^{\!1/2}
  \Ai(\hbar^{-2/3}\zeta)^2+\varepsilon_2(\hbar,x).
\end{displaymath}
It is known that
\begin{displaymath}
  \Ai(x) \leq \frac{1}{2\sqrt{\pi}}\,x^{-1/4}
  \exp\!\left(-\frac{2}{3}\,\hbar^{-1}x^{3/2}\right)
  \quad\textrm{for~}x>0,
\end{displaymath}
see \cite[Chapter~11]{olver74:_asymptotics_special_funct}. Using also
the estimates of error terms from Theorem~\ref{thm:olver_1turn_pt} one
can check that
\begin{displaymath}
  |\varepsilon_2(\hbar,x)| \leq \const\,f^{-1/2}
  \exp\!\left(-\frac{4}{3}\,\hbar^{-1}\zeta^{3/2}\right)\hbar^{4/3}
  \quad{for~}x>x_+.
\end{displaymath}
It follows that 
\begin{eqnarray}
  \left|\int_{x_+}^\infty\varepsilon_2(\hbar,x)\,\dd x\right|
  &\leq& \const\,\hbar^{4/3}\int_{x_+}^\infty f^{-1/2}
  \exp\!\left(-\frac{4}{3}\,\hbar^{-1}\zeta^{3/2}\right)\dd x
  \nonumber\\
  \label{eq:int_e2_estim}
  &=& \const\,\hbar^{4/3}\int_0^\infty
  \frac{y^{1/2}}{f\bigl(\xi(y)\bigr)}\,
  \exp\!\left(-\frac{4}{3}\,\hbar^{-1}y^{3/2}\right)\dd y.
\end{eqnarray}
There exists $c\geq0$ such that for $y>0$,
$f\bigl(\xi(y)\bigr)^{-1}\leq{}c(1+y^{-1})$. Hence
(\ref{eq:int_e2_estim}) is majorized by
\begin{displaymath}
  \const\,\hbar^{4/3}\int_0^\infty(y^{1/2}+y^{-1/2})
  \exp\!\left(-\frac{4}{3}\,\hbar^{-1}y^{3/2}\right)\dd y
  = O(\hbar^{5/3}).
\end{displaymath}

The asymptotic formula (\ref{eq:asympt_Ai_minf}) implies that
$|\Ai(x)|\leq\const\,|x|^{-1/4}$ for $x<0$. Recalling once more
Theorem~\ref{thm:olver_1turn_pt} we have
\begin{equation}
  \label{eq:int_e2_x1_xp}
  \left|\int_{x_1}^{x_+}\varepsilon_2(\hbar,x)\,\dd x\right|
  \leq \const\,\hbar^{4/3}\int_{x_1}^{x_+}|f|^{-1/2}\,\dd x
  = O(\hbar^{4/3}).
\end{equation}
This concludes the verification of (\ref{eq:int_psi2_infty}).

\noindent\addtocounter{point}{1}\textit{(\thepoint)~Oscillating integral
  on the interval $]x_1,x_+-\delta[$}.
By the usual integration by parts one can verify the following claim.

\noindent\textit{Let $[a,b]$ be a compact interval,
  $F\in{}C^1([a,b])$, $\mu\in{}C^2([a,b])$ and $\nu(\hbar,z)$ be twice
  continuously differentiable in $z$ on $[a,b]$. Assume that $\mu'(z)$
  nowhere vanishes on $[a,b]$ and}
\begin{displaymath}
  \sup_{z\in[a,b]}|\partial_z\nu(\hbar,z)| = O(1),\textrm{~}
  \sup_{z\in[a,b]}|\partial_z^2\nu(\hbar,z)| = O(1).
\end{displaymath}
\textit{Then for all sufficiently small $\hbar$ it holds true that}
\begin{displaymath}
  \left|\int_a^b F(z)
    \sin\bigl(\hbar^{-1}\mu(z)+\nu(\hbar,z)\bigr)\,\dd z\right|
  \leq \const\,\hbar
\end{displaymath}
\textit{where the constant depends only on the length of the interval
  $[a,b]$ and on the quantities}
\begin{displaymath}
  \mu_0^{\,-1}\|F\|_C,\textrm{~}
  \mu_0^{\,-2}\|F\|_C\|\mu''\|_C,\textrm{~}\mu_0^{\,-1}\|F'\|_C,
\end{displaymath}
\textit{with}
\begin{displaymath}
  \mu_0 = \min_{z\in[a,b]}|\mu'(z)|
\end{displaymath}
\textit{and $\|\cdot\|_C$ standing for the norm in the Banach space
  $C([a,b])$.}

As a consequence we find that if $W\in{}C^1(\R)$ then
\begin{equation}
  \label{eq:asympt_W_osc}
  \int_{x_1}^{x_+-\delta}\frac{W}{\sqrt{E-V}}\,
  \sin\!\left(\frac{2}{3}\,\hbar^{-1}
    \bigl(|\zeta|^{3/2}+|\tilde{\zeta}|^{3/2}\bigr)\right)\dd x
  = \delta^{-1}O(\hbar).
\end{equation}
To show this asymptotics it suffices to set in the above claim
$F=W/\sqrt{E-V}$, $\mu=(4/3)|\zeta|^{3/2}$ and
\begin{eqnarray*}
  \nu(\hbar,z) &=& \frac{2}{3}\,\hbar^{-1}
  \bigl(|\tilde{\zeta}(z)|^{3/2}-|\zeta(z)|^{3/2}\bigr)\\
  &=& \hbar^{-1}\left(\int_{z}^{\tilde{x}_+}
    \sqrt{\tilde{E}-V(t)}\,\dd t
    -\int_z^{x_+}\sqrt{E-V(t)}\,\dd t\right).
\end{eqnarray*}
Hence $\mu'(z)=-2\sqrt{E-V(z)}$ and
\begin{eqnarray*}
  \partial_z\nu(\hbar,z) &=& \frac{E-\tilde{E}}{\hbar}\,
  \left(\sqrt{E-V(z)}+\sqrt{\tilde{E}-V(z)}\,\right)^{\!-1},\\
  \partial_z^2\nu(\hbar,z) &=& \frac{E-\tilde{E}}{2\hbar}\,
  V'(z)\bigl(E-V(z)\bigr)^{-1/2}\bigl(\tilde{E}-V(z)\bigr)^{-1/2}\\
  && \times\left(\sqrt{E-V(z)}+\sqrt{\tilde{E}-V(z)}\,\right)^{\!-1}.
\end{eqnarray*}

\noindent\addtocounter{point}{1}\textit{(\thepoint)~The leading
  asymptotic term on the interval $]x_1,x_+-\delta[$}.
Let us check the contribution to the matrix element coming from the
interval $[x_1,x_+-\delta]$. The leading asymptotic term in the
expansion of $\psi$ is given in (\ref{eq:olver_psi_Ai}). We also need
the asymptotic behavior of the Airy function (\ref{eq:asympt_Ai_minf})
and the fact that the function $f/\zeta$ is continuous and hence
bounded on the interval $[x_1,x_+]$. We conclude that
\begin{displaymath}
  \psi \sim \left(\frac{\zeta}{f}\right)^{\!1/4}\Ai(\hbar^{-2/3}\zeta)
  = \frac{\hbar^{1/6}}{\sqrt{\pi}\,|f|^{1/4}}\,
  \cos\!\left(\frac{2}{3}\,\hbar^{-1}|\zeta|^{3/2}
    -\frac{\pi}{4}\right)+\frac{1}{|f|^{7/4}}\,O(\hbar^{7/6}).
\end{displaymath}
Observe that
\begin{displaymath}
  \hbar^{4/3}\int_{x_1}^{x_+-\delta}\frac{\dd x}{|f|^2}
  = \delta^{-1}O(\hbar^{4/3}),
\end{displaymath}
and on the interval $[x_1,x_+-\delta]$,
\begin{displaymath}
  (\tilde{E}-V)^{-1/4}
  = (E-V)^{-1/4}\bigl(1+\delta^{-1}O(\hbar)\bigr).
\end{displaymath}
From the boundedness of $W$ and from an estimate similar to
(\ref{eq:int_e2_x1_xp}) it follows that
\begin{eqnarray*}
  \int_{x_1}^{x_+-\delta}W\psi\tilde{\psi}\,\dd x
  &=& \int_{x_1}^{x_+-\delta}W\left(\frac{\zeta}{f}\right)^{\!1/4}
  \left(\frac{\tilde\zeta}{f}\right)^{\!1/4}
  \Ai(\hbar^{-2/3}\zeta)\Ai(\hbar^{-2/3}\tilde\zeta)\,\dd x
  + O(\hbar^{4/3}) \\
  &=& \frac{\hbar^{1/3}}{\pi}\int_{x_1}^{x_+-\delta}
  \frac{W}{|f|^{1/2}}\,\bigl(1+\delta^{-1}O(\hbar)\bigr) \\
  && \quad\times \cos\!\left(
    \frac{2}{3}\,\hbar^{-1}|\zeta|^{3/2}-\frac{\pi}{4}\right)
  \cos\!\left(\frac{2}{3}\,\hbar^{-1}|\tilde\zeta|^{3/2}
    -\frac{\pi}{4}\right)\dd x \\
  && +\,\delta^{-1}O(\hbar^{4/3}).
\end{eqnarray*}
Using the asymptotic behavior (\ref{eq:asympt_W_osc}) we have
\begin{equation}
  \label{eq:Wpsitpsi_cos_zeta}
  \int_{x_1}^{x_+-\delta}W\psi\tilde{\psi}\,\dd x
  = \frac{\hbar^{1/3}}{2\pi}\int_{x_1}^{x_+-\delta}
  \frac{W}{\sqrt{E-V}}\,
  \cos\!\left(\frac{2}{3}\,\hbar^{-1}
    (|\zeta|^{3/2}-|\tilde\zeta|^{3/2})\right)\dd x
  +\delta^{-1}O(\hbar^{4/3}).
\end{equation}

\noindent\addtocounter{point}{1}\textit{(\thepoint)~The argument of
  the cosine on the interval $]x_1,x_+-\delta[$}.
Let us show that for $x\in[x_1,x_+-\delta]$,
\begin{equation}
  \label{eq:zeta-tzeta_asympt}
  \frac{2}{3}\,\hbar^{-1}(|\zeta|^{3/2}-|\tilde\zeta|^{3/2})
  = -\frac{2\pi k}{T}\,\tau(x)+\delta^{1/2}O(1)
\end{equation}
where $\tau(x)$ was defined in (\ref{eq:tau_def}). We have
\begin{eqnarray*}
  \frac{2}{3}\,\hbar^{-1}(|\zeta|^{3/2}-|\tilde\zeta|^{3/2})
  &=& \hbar^{-1}\left(\int_x^{x_+}\sqrt{E-V}\,\dd t
    -\int_x^{\tilde{x}_+}\sqrt{\tilde{E}-V}\,\dd t\right) \\
  &=& \hbar^{-1}\bigg(\int_x^{x_+-\delta}
    \left(\sqrt{E-V}-\sqrt{\tilde{E}-V}\right)\dd t \\
    &&\qquad +\int_{x_+-\delta}^{x_+}\sqrt{E-V}\,\dd t
    -\int_{x_+-\delta}^{\tilde{x}_+}\sqrt{\tilde{E}-V}\,\dd t\bigg).
\end{eqnarray*}
Set temporarily
\begin{displaymath}
  g(y) = \int_{x_+-\delta}^y \sqrt{V(y)-V(t)}\,\dd t.
\end{displaymath}
Then for $y$ lying between $x_+$ and $\tilde{x}_+$ it holds
\begin{eqnarray*}
  |g'(y)| &=& \left|\frac{1}{2}\int_{x_+-\delta}^y
    \frac{V'(y)}{\sqrt{V(y)-V(t)}}\,\dd t\right|
  \leq \frac{1}{2}\,\const\int_{x_+-\delta}^y
  \frac{\dd t}{\sqrt{y-t}} \\
  &\leq& \const\,\sqrt{|x_+-\tilde{x}_+|+\delta}.
\end{eqnarray*}
Hence
\begin{eqnarray}
  \left|\int_{x_+-\delta}^{x_+}\sqrt{E-V}\,\dd t
  -\int_{x_+-\delta}^{\tilde{x}_+}\sqrt{\tilde{E}-V}\,\dd t\right|
  &=& |g(x_+)-g(\tilde{x}_+)| \nonumber\\
  &\leq& \const\,\sqrt{|x_+-\tilde{x}_+|+\delta}\,|x_+-\tilde{x}_+| 
  \nonumber\\
  \label{eq:g-tg_asympt}
  &=& \delta^{1/2}O(\hbar).
\end{eqnarray}
Furthermore,
\begin{eqnarray*}
  \sqrt{E-V}-\sqrt{\tilde{E}-V}-\frac{E-\tilde{E}}{2\sqrt{E-V}}
  &=& \frac{(E-\tilde{E})^2}{2
    \left(\sqrt{E-V}+\sqrt{\tilde{E}-V}\right)^{\!2}\sqrt{E-V}} \\
  &\leq& \frac{(E-\tilde{E})^2}{2(E-V)^{3/2}}
\end{eqnarray*}
and
\begin{displaymath}
  \int_x^{x_+-\delta}\frac{(E-\tilde{E})^2}{(E-V)^{3/2}}\,\dd t
  = \delta^{-1/2}O(\hbar^2).
\end{displaymath}
From (\ref{eq:Eprim-E_eq_Oh}) it follows that
\begin{displaymath}
  \hbar^{-1}(\tilde{E}-E) = \frac{2\pi k}{T}+O(\hbar)
\end{displaymath}
where $T$ is the period of the classical motion, see
(\ref{eq:T_def}). Altogether this means that
\begin{eqnarray}
  \hbar^{-1}\int_x^{x_+-\delta}
  \left(\sqrt{E-V}-\sqrt{\tilde{E}-V}\right)\dd t
  &=& -\left(\frac{2\pi k}{T}+O(\hbar)\right)
  \int_x^{x_+-\delta}\frac{\dd t}{2\sqrt{E-V(t)}} \nonumber\\
  && +\, \delta^{-1/2}O(\hbar) \nonumber\\
  \label{eq:int_sqrt_E-V_asympt}
  &=& -\frac{2\pi k}{T}\int_x^{x_+}\frac{\dd t}{2\sqrt{E-V(t)}} \\
  && +\,\delta^{1/2}O(1)+\delta^{-1/2}O(\hbar). \nonumber
\end{eqnarray}
Relations (\ref{eq:g-tg_asympt}) and (\ref{eq:int_sqrt_E-V_asympt})
jointly imply (\ref{eq:zeta-tzeta_asympt}).

\noindent\addtocounter{point}{1}\textit{(\thepoint)~The final step}.
From (\ref{eq:Wpsitpsi_cos_zeta}) and (\ref{eq:zeta-tzeta_asympt}) we
derive that
\begin{eqnarray}
  \int_{x_1}^{x_+-\delta}W\psi\tilde{\psi}\,\dd x
  &=& \frac{\hbar^{1/3}}{2\pi}\bigg(\int_{x_1}^{x_+-\delta}
  \frac{W(x)}{\sqrt{E-V(x)}}\, \nonumber\\
  &&\qquad \times\cos\!\left(\frac{2\pi k}{T}\,\tau(x)
    +\delta^{1/2}O(1)+\delta^{-1/2}O(\hbar)\right)\dd x
  +\delta^{-1}O(\hbar)\bigg) \nonumber\\
  \label{eq:Wpsitpsi_cos}
  &=& \frac{\hbar^{1/3}}{2\pi}\left(\int_{x_1}^{x_+}
  \frac{W(x)}{\sqrt{E-V(x)}}\,
  \cos\!\left(\frac{2\pi k}{T}\,\tau(x)\right)\dd x
  +\delta^{1/2}O(1)\right).
\end{eqnarray}
The interval $[x_-+\delta,x_1]$ can be treated similarly. We have
\begin{displaymath}
  \int_{x_-+\delta}^{x_1}W\psi\tilde{\psi}\,\dd x
  = \kappa\tilde{\kappa}\,
  \frac{\hbar^{1/3}}{2\pi}\left(\int_{x_-}^{x_1}
    \frac{W(x)}{\sqrt{E-V(x)}}\,
    \cos\!\left(\frac{2\pi k}{T}\,\tau_-(x)\right)\dd x
    +\delta^{1/2}O(1)\right)
\end{displaymath}
where
\begin{displaymath}
  \tau_-(x) = \frac{1}{2}\int_{x_-}^x\frac{\dd y}{\sqrt{E-V(y)}}
  = \frac{1}{2}\,T-\tau(x).
\end{displaymath}
Taking into account also (\ref{eq:kappa_asympt}) and
(\ref{eq:tkappa_asympt}) we finally find that
\begin{equation}
  \label{eq:Wpsitpsi_cos_minus}
  \int_{x_-+\delta}^{x_1}W\psi\tilde{\psi}\,\dd x
  = \frac{\hbar^{1/3}}{2\pi}\left(\int_{x_-}^{x_1}
    \frac{W(x)}{\sqrt{E-V(x)}}\,
    \cos\!\left(\frac{2\pi k}{T}\,\tau(x)\right)\dd x
    +\delta^{1/2}O(1)\right).
\end{equation}

From the boundedness of $W$ and relations (\ref{eq:int_psi2_infty}),
(\ref{eq:Wpsitpsi_cos}) and (\ref{eq:Wpsitpsi_cos_minus}) it follows
that
\begin{equation}
  \label{eq:int_R_Wpsitpsi}
  \int_\R W\psi\tilde{\psi}\,\dd x
  = \frac{\hbar^{1/3}}{2\pi}\left(\int_0^T W\bigl(q(t)\bigr)
  e^{ik\omega t}\,\dd t+\delta^{1/2}O(1)\right).
\end{equation}
As a particular case, with $W(x)=1$ and $k=0$, we have
\begin{equation}
  \label{eq:int_R_psi2}
  \int_\R \psi^2\,\dd x
  = \frac{\hbar^{1/3}}{2\pi}\big(T+\delta^{1/2}O(1)\big).
\end{equation}
The same relation holds also for the squared norm of $\tilde\psi$.

Relations (\ref{eq:int_R_Wpsitpsi}) and (\ref{eq:int_R_psi2}) imply
that there exists $c\geq0$ such that for all sufficiently small
positive $\delta$ and all $n$, $n\geq{}n_0(\delta)$, it holds
\begin{displaymath}
  \left|\langle{}n|W(x)|n+k\rangle-\frac{1}{T}
    \int_0^T W\!\left(q(t)\right)e^{ik\omega t}\dd t\right|
  \leq c\delta^{1/2}.
\end{displaymath}
Since $\delta$ is arbitrary this concludes the verification of the
limit (\ref{eq:nWnk_to_semic}).

\section*{Acknowledgments}
The authors wish to acknowledge gratefully partial support from the
following grants: grant No.~201/05/0857 of the Grant Agency of the
Czech Republic (P.~\v{S}.), grant No.~MSM 6840770010 of the Ministry
of Education of the Czech Republic (O.~L.), and grant No.~LC06002 of
the Ministry of Education of the Czech Republic (the both authors).

%\bibliographystyle{/data/work/clanky/bibtex/myplain.bst}
%\bibliography{/data/work/clanky/bibtex/mojeref.bib}%,%
%%/data/work/clanky/bibtex/psref.bib}
%\end{document}

\end{document}